\documentclass[journal]{IEEEtran}
\usepackage{eso-pic}
\usepackage{cite}
\usepackage{orcidlink}
\usepackage{blindtext}
\usepackage{verbatim}

\usepackage{graphicx}
\usepackage{tabularx}
\usepackage{silence}
\usepackage[labelformat=simple]{subcaption}
\usepackage{pgfplots}
\usepgfplotslibrary{groupplots}
\usepackage{tikz}
\usetikzlibrary{patterns}
\usetikzlibrary{spy}
\pgfplotsset{compat=1.18}

\definecolor{tu_1}{RGB}{0,90,169}
\definecolor{tu_2}{RGB}{0,131,204}
\definecolor{tu_3}{RGB}{0,157,129}
\definecolor{tu_4}{RGB}{153,192,0}
\definecolor{tu_5}{RGB}{201,212,0}
\definecolor{tu_6}{RGB}{253,202,0}
\definecolor{tu_7}{RGB}{245,163,0}
\definecolor{tu_8}{RGB}{236,101,0}
\definecolor{tu_9}{RGB}{230,0,26}
\definecolor{tu_10}{RGB}{166,0,132}
\definecolor{tu_11}{RGB}{114,16,133}

\usepackage{amsmath}
\usepackage{amssymb}
\usepackage{siunitx}
\usepackage{numprint}
\npdecimalsign{.}
\nprounddigits{2}

\renewcommand{\vec}[1]{\mathbf{#1}}

\newcommand{\Jc}{j_{\text{c}}}

\newcommand{\h}{\vec{h}}
\newcommand{\ephi}{\hat{\vec e}_{\varphi}}

\newcommand{\omegac}{\Omega_{\text{c}}}

\newcommand{\surfInt}[2]{\int_{#2} #1 \ \mathrm{d}#2}

\newcommand{\Lzero}{L'_0}
\newcommand{\Flux}{\Phi}
\newcommand{\Fluxint}{\Flux'_{\text{int}}}
\newcommand{\Fluxintk}[1]{\Flux'_{\text{int, #1}}}
\newcommand{\Fluxintdt}{\Dot{\Flux}'_{\text{int}}}

\newcommand{\V}{V'}
\newcommand{\Vc}{\V_{\text{c}}}
\newcommand{\I}{I}
\newcommand{\Ic}{\I_{\text{c}}}

\newcommand{\Irev}{\I_{\text{rev}}}
\newcommand{\Irevdt}{\Dot{\I}_{\text{rev}}}
\newcommand{\Irevk}[1]{\I_{\text{rev}, #1}}
\newcommand{\Irevkdt}[1]{\Dot{\I}_{\text{rev}, #1}}
\newcommand{\Iirr}{\I_{\text{irr}}}
\newcommand{\Iirrk}[1]{\I_{\text{irr}, #1}}
\newcommand{\Ieddy}{\I_{\text{eddy}}}
\newcommand{\Ieddyk}[1]{\I_{\text{eddy}, #1}}
\newcommand{\G}{G}
\newcommand{\Gprev}{\G^{\text{prev}}}

\newcommand{\Weight}{\beta}
\newcommand{\Coerc}{\xi}
\newcommand{\Timec}{\eta}
\newcommand{\Weightk}[1]{\Weight_{#1}}
\newcommand{\Coerck}[1]{\Coerc_{#1}}
\newcommand{\Timeck}[1]{\Timec_{#1}}
\newcommand{\Coercf}{f_\Coerc(b, T)}

\renewcommand{\P}{P'}

\newcommand{\Pirr}{\P_{\text{irr}}}
\newcommand{\Peddy}{\P_{\text{eddy}}}
\newcommand{\Pjoule}{\P_{\text{joule}}}
\newcommand{\Ptot}{\P_{\text{total}}}

\newcommand{\R}{R'}
\newcommand{\Ifil}{\I_{\text{f}}}

\newcommand{\Rfil}{\R_{\text{f}}(\Ifil)}
\newcommand{\Rmat}{\R_{\text{m}}}
\newcommand{\Req}{\R_{\text{eq}}}

\newcommand{\Cellnum}{M}

\begin{document}

\title{Reduced Order Hysteretic Flux Model for Transport Current Homogenization in Composite Superconductors}
\author{Alexander~Glock$^\text{1,2}$~\orcidlink{0009-0009-7663-6471},
        Julien~Dular$^\text{1}$~\orcidlink{0000-0003-0503-7526},
        Arjan~Verweij$^\text{1}$~\orcidlink{0000-0001-6945-8608} and
        Mariusz~Wozniak$^\text{1}$~\orcidlink{0000-0002-3588-426X}
        \vspace{0.2cm}\\
        $~^\text{1}$ CERN, Geneva, Switzerland\\
        $~^\text{2}$ Technical University of Darmstadt, Darmstadt, Germany}
\markboth{Reduced Order Hysteretic Flux (ROHF) model, July~2025}%
{}
\maketitle

\begin{abstract}
In this paper, we present the Reduced Order Hysteretic Flux (ROHF) model to describe the relationship between time-varying transport current and internal magnetic flux for composite superconductors. The ROHF model is parametrized using reference simulations of the conductor response, after which it enables the computation of macroscopic quantities such as voltage and power loss without requiring detailed electromagnetic field solutions. It is therefore suitable for the homogenization of transport current effects in large-scale superconducting magnets, avoiding a fine discretization of the composite strand small-scale structures, allowing to drastically reduce the computational effort. The approximation can be implemented either (i) as a rate-independent model, neglecting eddy current effects in the normal conducting matrix, or (ii) as a rate-dependent model, including those eddy current effects. In this paper, the complete modeling workflow, including parameter identification, coupling with the other fields (temperature and magnetic field), and post-processing of the results, is described and verified using a twisted multifilamentary strand as an example.
\end{abstract}

\IEEEpeerreviewmaketitle

\AddToShipoutPicture*{
    \footnotesize\sffamily\raisebox{0.7cm}{\hspace{1.7cm}\fbox{
        \parbox{0.97\textwidth}{
            This work has been submitted to a journal for possible publication. Copyright may be transferred without notice, after which this version may no longer be accessible.
            }
        }
    }
}

\section{Introduction}

\begin{figure}[!ht]
\centering
\includegraphics[width=\linewidth]{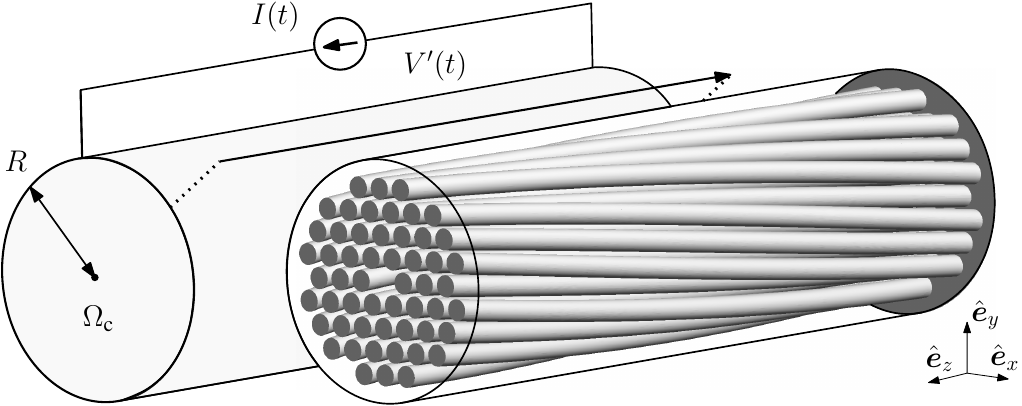}
\caption{Fully resolved model of a twisted multifilamentary superconductor strand (right) and the equivalent homogenized model without fine structure resolution (left).}
\label{fig:strand_3D}
\end{figure}

\IEEEPARstart{D}{eveloping} quench protection systems for superconducting magnets requires accurate simulations of their transient physical behavior, which are computationally demanding due to strong material nonlinearities, multi-scale geometries, and the coupled electromagnetic, thermal, and mechanical nature of the problem~\cite{Ravaioli25, Mulder2023}. The computation time for such multiphysics Finite Element (FE) simulations can be significantly reduced through homogenization methods, which abstract fine structures into regions with equivalent macroscopic characteristics~\cite{Feddi1997}. One of the challenges with this process lies in the accurate description of the current density to voltage relationship within composite superconductor strands. These strands typically consist of a high number of twisted superconducting filaments (e.g., Nb-Ti, $\text{Nb}_3$Sn, or MgB$_2$) embedded in a stabilizing normal-conducting matrix (Fig.~\ref{fig:strand_3D} - right). A homogenized model of equal dimensions (Fig.~\ref{fig:strand_3D} - left) does not fully resolve the fine structures, but rather relies on an averaged material relationship linking the applied transport current $\I$ ($\si{\ampere}$) to the voltage per unit length $\V$ ($\si{\volt/\meter}$) and the resulting power loss within the volume.

Homogenization methods for non-superconducting electromagnetic systems have been extensively studied~\cite{Sabariego2008}. However, directly applying them to superconducting systems is not possible as they exhibit a hysteretic response. In this work, we propose the novel Reduced Order Hysteretic Flux (ROHF) model for transport current effects homogenization in composite strands without detailed field solutions, utilizing a thermodynamically consistent hysteresis model developed for magnetization in ferromagnetic materials~\cite{bergqvist1997,Henrotte06} and recently adapted to model magnetization of composite superconductors~\cite{dular25rohm}. Reference numerical solutions are used to tune the ROHF model parameters. They are obtained using a detailed FE model, which can be efficiently implemented for periodic structures with the Coupled Axial and Transverse currents (CATI) method~\cite{dular24cati}.

\section{Internal magnetic flux in composite strands} 

\begin{figure}[ht!]
\centering
\begin{subfigure}[t]{0.99\linewidth}  
\centering
\begin{tikzpicture}[trim axis right, trim axis left][font=\small]
\pgfplotsset{set layers}
 	\begin{axis}[
	tick label style={/pgf/number format/fixed},
    width=\linewidth,
    height=3.8cm,
    ylabel={$\Fluxint$ (Wb/m)},
    ytick={yshift=-0.5em},
    yticklabel style={xshift=0em},
    axis line style={draw=none},
    tick style={draw=none},
    xticklabel=\empty,
    yticklabel=\empty,
    yticklabel pos=left,
    ]
    \end{axis}
 	\begin{axis}[
	tick label style={/pgf/number format/fixed},
    width=\linewidth,
    height=4.5cm,
    grid = major,
    grid style = dotted,
    ymin=-5.5e-4, 
    ymax=5.5e-4,
    xtick={-2960, 0, 2960},
    xticklabels={$-\Ic$, 0 , $\Ic$},
	xlabel={Transport current},
    ylabel style={yshift=-0.5em},
    xlabel style={yshift=0.5em},
    xticklabel style={yshift=0.1em},
    yticklabel pos=right,
    legend columns=2,
    legend style={at={(0.21, 0.6)}, cells={anchor=west}, anchor=south, draw=none,fill opacity=0, text opacity = 1, legend image code/.code={\draw[##1,line width=1pt] plot coordinates {(0cm,0cm) (0.3cm,0cm)};}}
    ]
    \addplot[tu_3, thick] 
    table[x=f0.1_IIC1.0_x,y=f0.1_IIC1.0_y]{data/current_flux_ratedep.txt};
    \addplot[tu_6, thick] 
    table[x=f10_IIC1.0_x,y=f10_IIC1.0_y]{data/current_flux_ratedep.txt};
    \addplot[tu_9, thick] 
    table[x=f1000_IIC1.0_x,y=f1000_IIC1.0_y]{data/current_flux_ratedep.txt};
    \legend{$0.1 \si{\hertz}$, $10 \si{\hertz}$, $1 \si{\kilo\hertz}$};
    \end{axis}
\end{tikzpicture}
\end{subfigure}
\caption{Internal flux hysteresis in a superconducting strand according to full scale FE references for three frequencies of sinusoidal transport current with the critical current amplitude $\Ic$.}
\label{fig:hysteresis}
\end{figure}
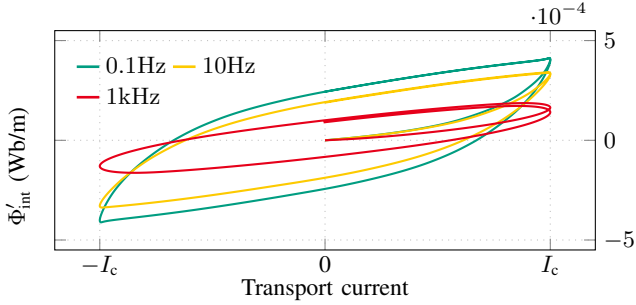

Injecting a transient transport current $\I(t)$ in a composite superconducting strand leads to radial inward penetration of current into the material with accompanying dissipation of energy due to the shifting of magnetic flux vortices. Saturation occurs at the critical transport current $\Ic$, when the entire strand cross section carries the critical current density $\Jc$. During this process, natural or artificial imperfections in the material pin the flux vortices, creating local energy barriers that must be overcome for flux motion~\cite{geneko1994,dular25rohm}. This flux pinning leads to a hysteretic relationship between the internal magnetic flux per unit length, $\Fluxint$ ($\si{\weber/\meter}$), and changes in the transport current. The hysteretic internal flux gives rise to a history-dependent voltage per unit length
\begin{equation}\label{eqn:transport_voltage}
\begin{aligned}
V'(t) = - \Fluxintdt
\end{aligned}
\end{equation}
in $\si{\volt/\meter}$, where $\Fluxintdt$ denotes the time derivative of $\Fluxint$. This voltage does not scale linearly with the transport current rate of change and can therefore not be calculated with a constant inductance. In order to reproduce the voltage and the related loss, it is necessary to model the internal magnetic flux as a function of the time-varying transport current $\I(t)$. For composite superconductors with an approximately rotational symmetric cross section $\omegac$, introducing cylindrical coordinates $(r,\phi, z)$, the internal flux per unit length can be defined as follows:
\begin{align}\label{eqn:Fluxint_integral}
\Fluxint &= \surfInt{\frac{1}{2\pi r} \ \mu_0\,\h \cdot \ephi}{\omegac},
\end{align}
with the magnetic permeability of vacuum $\mu_0 = 4\pi \times 10^{-7}$ $\si{\henry/\meter}$, and $\h$ the magnetic field ($\si{\ampere/\meter}$). Equation~\eqref{eqn:Fluxint_integral} can be interpreted as the average of the magnetic fluxes across radial paths of the strand cross-section in all azimuthal directions. If subjected to sinusoidal transport current, such a strand generates a hysteresis loop of internal flux (Fig.~\ref{fig:hysteresis}) whose form is not only dependent on the applied current amplitude but also on its rate of change. With pure transport current excitations, inter-filament coupling currents can be neglected.

Extracting the instantaneous power loss from the internal flux hysteresis curve is not directly possible, as the change in internal flux not only contains dissipation due to flux shifting but also contains reversible magnetic energy exchanges. For periodic transport current and voltage signals, the dissipated power loss can be recovered by integrating $\I(t)\V(t)$ over a full signal period, during which the energy stored and retrieved from the magnetic field cancels out. The ROHF model, presented in the next section, lifts this limitation of the power loss computation to periodic excitations by separating stored and dissipated energy at any time instant.

\section{Reduced Order Hysteretic Flux (ROHF) model} 

\begin{figure}[ht]
\centering
\includegraphics[width=0.9\linewidth]{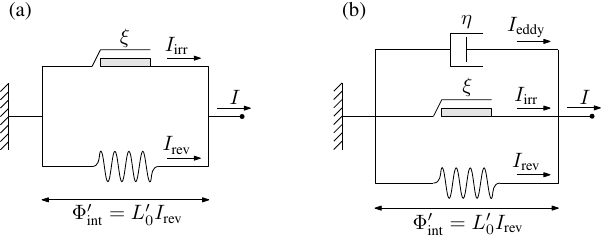}
\caption{Mechanical analogy of the (a) rate-independent and (b) rate-dependent ROHF cell.}
\vspace{-0.5cm}
\label{fig:rohf_cell}
\end{figure}

In order to reproduce the hysteretic relationship between the internal flux $\Fluxint$ and the transport current $I(t)$ without a need to compute the detailed current density distribution, we propose to adapt the energy-based hysteresis model, originally developed for the description of ferromagnetic hysteresis \cite{bergqvist1997,Henrotte06}. This model can be efficiently implemented by means of reduced order cells~\cite{Henrotte06} and was recently adapted to superconducting strands for the homogenization of magnetization effects~\cite{dular25rohm}.
A mechanical analogy for the proposed adaptation as Reduced Order Hysteretic Flux (ROHF) cell for a (a) rate-independent model, as well as for its (b) rate-dependent extension is that of a massless element subjected to dry friction, a linear spring, and a linear damper~\cite{bergqvist1997,Henrotte2006dyn,glock25}, as illustrated in Fig.~\ref{fig:rohf_cell}.

\subsection{Rate-independent ROHF cell} 

In both implementations of the ROHF cell, the transport current injected into the strand acts comparable to the driving force in the mechanical analogy. In the rate-independent cell shown in Fig.~\ref{fig:rohf_cell}(a), this total transport current will be split into a reversible and an irreversible component:
\begin{equation}\label{eqn:current_split}
\begin{aligned}
\I = \Irev + \Iirr.
\end{aligned}
\end{equation}
Each current contribution is linked to one physical mechanism, clearly separating the energy within the system at each time instant. The reversible current $\Irev$ feeds the magnetic field and is therefore linked to the stored magnetic energy, which can be recuperated when reversing the transport current. Therefore, $\Irev$ drives the internal magnetic flux   
\begin{equation}\label{eqn:Fluxint}
\begin{aligned}
\Fluxint = \Weightk \ \Lzero \ \Irev,
\end{aligned}
\end{equation}
through a weight parameter $\Weight$ and a normalization constant $\Lzero = \frac{\mu_0}{4 \pi}$ ($\si{\henry/\meter}$), which is equivalent to the unlinked, internal self-inductance of a cross-sectional element of a cylindrical conductor in open space, assuming linear material with a uniform current density distribution \cite{Clayton+2009}. The apostrophes in Eq.~\eqref{eqn:Fluxint} indicate the normalization of quantities per unit length in the axial direction.

The irreversible current $\Iirr$ drives the dissipative flux shifting mechanism modeled through a dry-friction-like element. When the magnitude of the irreversible current exceeds a defined coercivity threshold $\Coerc$, the transport current supplies enough energy to overcome the pinning force on flux vortices. This results in power loss that is proportional to the rate of change of the excitation. The behavior is captured by the following cell update equation:
\begin{align}\label{eqn:scalar_play}
\Irev &= \left\{\begin{aligned}
& \Irev^\text{prev} \quad &&\text{if } \left| \I - \Irev \right| < \Coerc,\\
& \I  - \Coerc \ \dot \I /|\dot \I |\quad &&\text{if } \left|\I  - \Irev\right| \geq \Coerc,\\
\end{aligned}\right.
\end{align}
utilizing previously computed quantitites with superscript “$\text{prev}$" and the total transport current $\I$.

Combining Eqs.~\eqref{eqn:transport_voltage}, \eqref{eqn:current_split} and \eqref{eqn:Fluxint} the instantaneous hysteretic power loss of the ROHF cell is calculated as
\begin{equation}\label{eqn:Pirr}
\begin{aligned}
\Pirr = \Fluxintdt\ \Iirr = \Weight \ \Lzero \ \Irevdt \ \Iirr,
\end{aligned}
\end{equation}
based on the voltage per unit length along the conductor, Eq.~\eqref{eqn:transport_voltage}, given by the time derivative of Eq.~\eqref{eqn:Fluxint} and the irreversible current contribution. This power is equivalent to the dissipation within the superconducting filaments due to transport current transients.

A single cell yields a too simplistic approximation of the actual internal flux, and fails to describe the continuous internal flux hysteresis accurately. This problem is solved by combining multiple reduced order cells in a chain (described in Section~\ref{sec:chain_cells})~\cite{Henrotte06}.

\subsection{Rate-dependent ROHF cell} 

The previously derived reduced-order cell produces a rate-independent response. However, as noted earlier, the actual response of a composite strand is rate-dependent due to eddy currents (skin effect) in the normal-conducting matrix, while inter-filament coupling currents remain negligible. This rate dependence can be incorporated by introducing an additional contribution, $\Ieddy$, to the transport current decomposition~\cite{Henrotte2006dyn}, leading to:
\begin{equation}\label{eqn:current_split_eddy}
\begin{aligned} 
 \I = \Irev + \Iirr + \Ieddy,
\end{aligned}
\end{equation}
as illustrated in Fig.~\ref{fig:rohf_cell}(b), with $\Ieddy$ defined by
\begin{equation}\label{eqn:eddy}
\begin{aligned}
\Ieddy &= \Timec \ \Irevdt,
\end{aligned}
\end{equation}
where the dot notation indicates the time derivative of the reversible current $\Irev$ and $\Timec$ (s) is a time constant. The rate-dependent cell implementation is based on the lumped quantity $G = \Irev + \Ieddy$, which allows for a concise update equation for the rate-dependent cell:
\begin{align}\label{eqn:scalar_play_eddy}
\G &= \left\{\begin{aligned}
& \Gprev &&\text{if } \left|\I - \Gprev\right| < \Coerc,\\
& \I - \Coerc \ \dot \I/|\dot \I|\quad &&\text{if } \left|\I - \Gprev\right| \geq \Coerc,\\
\end{aligned}\right.
\end{align}
based on the value of the mixed quantity in the previous time step $\G^{\text{prev}}$. From Eq.~\eqref{eqn:scalar_play_eddy}, the lumping of reversible and eddy contributions within $G$ has to be resolved in an additional computation step, which is obtained from Eq.~\eqref{eqn:eddy} by approximating the first derivative of the reversible current with a backwards difference scheme~\cite{Henrotte2006dyn}. This leads to
\begin{equation}\label{eqn:Irev_approx}
\begin{aligned} 
\Irev &= \frac{1}{\Delta t+\Timec} \left( \Irev^{\textrm{prev}}\Timec + \G \ \Delta t  \right).
\end{aligned}
\end{equation}
The eddy current contribution can be approximated in a similar manner, yielding 
\begin{equation}\label{eqn:Ieddy_approx}
\begin{aligned} 
\Ieddy &= \frac{\Timec}{\Delta t+\Timec} \left(\G -  \Irev^{\textrm{prev}} \right).
\end{aligned}
\end{equation}
In practice, it is sufficient to compute and store the reversible current given by Eq.~\eqref{eqn:Irev_approx} as $\Ieddy$ can be obtained through Eq.~\eqref{eqn:eddy} with a finite difference approximation of the time derivative. With known eddy and reversible currents, the irreversible counterpart is given through Eq.~\eqref{eqn:current_split_eddy} for a total transport current $I$.

The eddy current associated power loss per unit length, dissipated in the matrix, can then be expressed as
\begin{equation}\label{eqn:Peddy}
\begin{aligned}
\Peddy = \Fluxintdt\, \Ieddy =  \Weight\, \Lzero\, \Irevdt\, \Ieddy = \Weight\, \Lzero\, \Timec\, \Irevdt^2.
\end{aligned}
\end{equation}
Eq.~\eqref{eqn:current_split_eddy}, is only an extension of the rate-independent case, Eq~\eqref{eqn:current_split}. Therefore the earlier derived expressions for hysteretic power loss per unit length, Eq.~\eqref{eqn:Pirr}, and internal flux per unit length, Eq.~\eqref{eqn:Fluxint}, of the rate-independent cell remain valid. As discussed in \cite{dular25rohm}, the modeling of eddy current effects with a linear first order ordinary differential equation is not accurate in the high frequency limit, as the actual eddy current induced power loss should decrease as $1/ \sqrt{f}$ for $f\to\infty$ due to the skin effect, while Eq.~\eqref{eqn:eddy} only leads to a decrease as $1/ f$ for $f\to\infty$. This inaccuracy for high frequency regimes is deemed acceptable as the model is intended to approximate transient power loss in superconducting magnets, that are typically operated at low to medium frequency well below $f <<  10~\si{\kilo\hertz}$.

\subsection{Magnetic field and temperature dependence} 

Until now, we neglected the influence of a magnetic background field and changing temperatures on internal flux hysteresis. However, those external effects have to be considered, since superconducting strands are usually supbject to significant background magnetic fields and experience local temperature changes due to power dissipation.

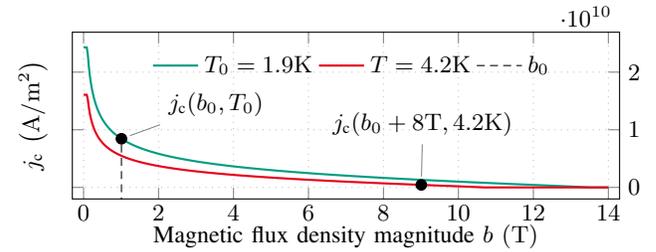
\begin{figure}[ht]
\centering
\begin{tikzpicture}[trim axis right, trim axis left][font=\small]
\pgfplotsset{set layers}
 	\begin{axis}[
	tick label style={/pgf/number format/fixed},
    width=\linewidth,
    height=3.8cm,
    ylabel={ $\Jc$ $\left(\si{\ampere}/\si{\meter}^2\right)$},
    ytick={yshift=-0.5em},
    yticklabel style={xshift=0em},
    axis line style={draw=none},
    tick style={draw=none},
    xticklabel=\empty,
    yticklabel=\empty,
    yticklabel pos=left,
    ]
    \end{axis}
 	\begin{axis}[
	tick label style={/pgf/number format/fixed},
    width=\linewidth,
    height=3.8cm,
    grid = major,
    grid style = dotted,
    xlabel={Magnetic flux density magnitude $b$ $\left(\si{\tesla}\right)$},
    xlabel style={yshift=0.5em},
    ymin=-0.2e10,
    xmin=-0.3,
    xmax=14.3,
    xticklabel style={yshift=0.1em},
    yticklabel style={xshift=0em},
    yticklabel pos=right,
    legend columns=3,
    legend style={at={(0.13, 0.8)}, cells={anchor=west}, anchor=west, draw=none,fill opacity=0, text opacity = 1}
    ]
    \addplot[color=tu_3, thick] 
    table[x=B,y=jc1.9]{data/jcb_nbti.txt};
    \addplot[color=tu_9, thick] 
    table[x=B,y=jc4.2]{data/jcb_nbti.txt};
    \addplot[color=black, densely dashed] table[row sep = crcr]{1.01 -1e10 \\ 1.01 0.844e10 \\};
    \legend{$T_0=1.9\si{\kelvin}$, $T=4.2\si{\kelvin}$, $b_0$};
    \addplot[mark=*] coordinates {(1.01, 0.844e10)} node[pin=20:{$\Jc(b_0,T_0)$}]{} ;
    \addplot[mark=*] coordinates {(9.01, 4.39e8)} node[pin=88:{$\Jc(b_0+8\si{\tesla},4.2\si{\kelvin})$}]{} ;
    \end{axis}
\end{tikzpicture}
\caption{Critical current density of Nb-Ti as a function of the magnetic flux density amplitude at two exemplary temperatures. The points indicate the base-line critical current density at ambient conditions $\Jc(b_0,T_0)$ and a deviating current density with $b\ne b_0$ and $T\ne T_0$ used to obtain a scaling factor.}
\label{fig:scaling}
\end{figure}

The ROHF model can be extended to account for those dependencies through the means of a field- and temperature-dependent coercivity
\begin{equation}\label{eqn:Coerc_scaling}
\begin{aligned}
\Coerc(b,T) = \Coercf \ \Coerc,
\end{aligned}
\end{equation}
where a scaling function $\Coercf$ dynamically adapts the coercivities to changes in magnetic field and temperature. Since the coercivities define transport current thresholds, the scaling function $\Coercf$ can be derived from the $\Jc(b,T)$ scaling of the superconductor used (see Fig.~\ref{fig:scaling}).

An effective coercivity scaling can therefore be implemented using the ratio between the critical current density at an arbitrary operating point $(b,T)$ and the initial critical current density $\Jc(b_0,T_0)$ — evaluated at the conductor cross-section average self-field magnitude of the strand and the base temperature used for the initial cell fitting
\begin{equation}\label{eqn:coerc_func}
\begin{aligned}
\Coercf = \frac{\Jc(b_0 + b,T)}{\Jc(b_0,T_0)},
\end{aligned}
\end{equation}
where $b$ represents the magnitude of an applied magnetic flux density perpendicular to the strand, and $T$ denotes the temperature. The average self-field amplitude $b_0$ during the initial chain fitting can be approximated by a constant or modeled via a dedicated self-field function $b_0(I)$ dependent on the transport current.

\subsection{Chain of cells}\label{sec:chain_cells} 

\begin{figure}[!ht]
\centering
\includegraphics[width=\linewidth]{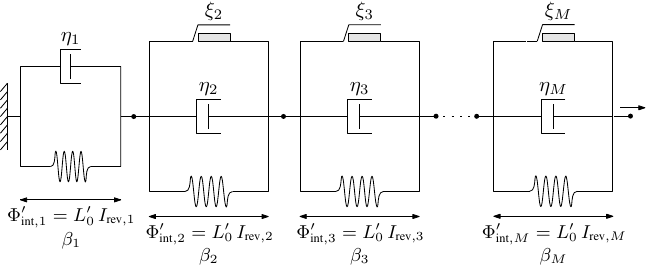}
\caption{Chain of $\Cellnum$ rate-dependent reduced order hysteretic flux cells.}
\label{fig:analogy_chain}
\end{figure}

Due to the fact that it is defined by a single threshold coercivity, the flux output of a single ROHF cell resembles a crude approximation of the complex hysteresis curves observed in composite strands (Fig.~\ref{fig:hysteresis}). In order to better represent the strand response, we can combine the output of multiple ROHF cells, as was proposed in~\cite{Henrotte06}. This leads to a chain configuration, as illustrated in Fig.~\ref{fig:analogy_chain}. In the chain, each cell is driven by the same transport current and contributes according to Eq.~\eqref{eqn:Fluxint} to the combined internal flux
\begin{equation}\label{eqn:Fluxint_chain}
\begin{aligned}
\Fluxint = \sum_{k=1}^\Cellnum  \Weightk{k} \ \Lzero \ \Irevk{k},
\end{aligned}
\end{equation}
where $\Cellnum$ is the number of cells and $\Weightk{k}$ is the inductance weight parameter and $\Irevk{k}$ the corresponding reversible current contribution of each cell. Spacing the set of cell coercivities $\Coerck{k}$ over a transport current interval yields a more sensitive system with gradual dissipation depending on the number of activated cells. The total hysteretic loss
\begin{equation}\label{eqn:Pirr_chain}
\begin{aligned}
\Pirr = \sum_{k=1}^\Cellnum  \Weightk{k} \ \Lzero \ \Irevkdt{k} \ \Iirrk{k},
\end{aligned}
\end{equation}
can be defined based on the voltage induced by the internal flux and the current split within each cell. Similarly, Eq.~\eqref{eqn:Peddy} for the eddy current induced loss within the normal conducting strand matrix can be extended for a multi-cell system:
\begin{equation}\label{eqn:Peddy_chain}
\begin{aligned}
\Peddy = \sum_{k=1}^\Cellnum  \Weightk{k} \, \Lzero \, \Irevkdt{k} \, \Ieddyk{k} = \sum_{k=1}^\Cellnum \Weightk{k}\, \Lzero\, \Timeck{k}\, \Irevkdt{k}^2,
\end{aligned}
\end{equation}
with the individual time constant $\Timeck{k}$ of each cell and $\Ieddyk{k}$ its respective eddy current contribution. The combination of $\Cellnum$ rate-dependent ROHF cells results in a total amount of $3\Cellnum$ free parameters, which have to be fitted to the individual composite superconducting strand. Strategies for this parameter identification are discussed in Section~\ref{sec:parameterIdentification}.

\subsection{Joule power loss} 

The hysteretic and eddy power loss described by Eqs.~\eqref{eqn:Pirr_chain} and \eqref{eqn:Peddy_chain} of the multi-cell ROHF model capture the two major loss contributions in an under-critical regime ($\I\leq\Ic$). However, when approaching the critical transport current of the strand, $\Ic$, Joule loss progressively becomes dominant until the complete transition of the material in to the normal state~\cite{campbell1982}. To capture this behavior, the ROHF model has to be extended by a transition to the over-critical regime including Joule power loss. In a first phenomenological approximation, this can be achieved through a current-dependent resistance
\begin{equation}\label{eqn:power_law}
\begin{aligned}
\Rfil = \frac{\Vc}{\Ic} \left(  \frac{|\Ifil|}{\Ic}\right)^{n-1},
\end{aligned}
\end{equation}
using a critical voltage per unit length $\Vc=10^{-4}\,\si{\volt/\meter}$. This power-law model~\cite{rhyner1993} approximates the strand resistance for transport current regimes below and up to the critical current level. However, it fails to account for the presence of the strand normal state resistance, which funnels part of the total transport current and thereby introduces a current-sharing effect.

This current sharing effect is included in an equivalent electric model, describing the parallel topology of the strand normal state resistivity $\Rmat$ and the power-law dependent resistance of the combined filaments $\Rfil$. The interdependence in this current sharing is resolved through an iterative scheme using the bisection-method~\cite{Bortot+2020}, solving for the equivalent total resistance
\begin{equation}\label{eqn:equi_resistance}
\begin{aligned}
\Req = \left(  \frac{1}{\Rmat} + \frac{1}{\Rfil}\right)^{-1}.
\end{aligned}
\end{equation}
In the overcritical regime $\I >> \Ic$ this equivalent resistance $\Req$ will approach the strand normal state resistance $\Rmat$, as the power law resistance of the quenched filaments becomes several orders of magnitude larger. Furthermore, the equivalent resistance $\Req$ allows for the computation of Joule power loss per meter
\begin{equation}\label{eqn:Pjoule}
\begin{aligned}
\Pjoule = \Req\, \I^2,
\end{aligned}
\end{equation}
which is added to the previously defined contributions of hysteretic and eddy power loss associated with the chain of rate-dependent ROHF cells given by Eqs.~\eqref{eqn:Pirr_chain} and \eqref{eqn:Peddy_chain} in order to obtain a total power loss approximation
\begin{equation}\label{eqn:Ptot}
\begin{aligned}
\Ptot = \Pirr + \Peddy + \Pjoule
\end{aligned}
\end{equation}
 of the ROHF model in $\si{\watt/\meter}$.

\section{Parameter identification}\label{sec:parameterIdentification} 

The ROHF model presented in the previous section was developed with the aim of replicating the internal magnetic flux of different geometries of composite superconducting strands when combined into a chain of $\Cellnum$ cells. The $3\Cellnum$ free parameters must be carefully selected to ensure accurate approximation of the reference solution; this process can be automated to avoid manual tuning, as outlined in this section.

The chosen approach for the automated parameter identification process relies on the error minimization between (i) the internal flux, Eq.~\eqref{eqn:Fluxint_integral}, on a grid of reference simulations spanning over ranges of transport current frequency and amplitude of a sinusoidal signal and (ii) the corresponding output of a ROHF chain, Eq.~\eqref{eqn:Fluxint_chain}, with a fixed number $\Cellnum$ of rate-dependent cells. The resulting optimization can be performed through a standard implementation of the limited-memory Broyden-Fletcher-Goldfarb-Shanno (L-BFGS) algorithm~\cite{Liu+1989} combining all free parameters in one optimization algorithm with the average, relative flux error over the first transport current cycle as metric. 
To facilitate convergence this optimization should be initialized with meaningful initial values for the cell parameters. The initial values of the $\Cellnum$ coercivities, $\Coerck{k}$,  can be defined as linearly spaced over the interval $[0, \Ic)$, excluding the critical current value, and enforcing $\Coerck{1} = 0$. A matching set of initial inductance weights for those initial coercivities can be obtained with the help of a high transport current amplitude reference simulation, which provides internal flux values up to strand saturation at a low transport current frequency such that rate-dependent effects can be neglected. This enables the following recursion scheme for initial inductance weights
\begin{equation}\label{eqn:recursion}
\begin{aligned}
\Weightk{1} = \frac{\Fluxintk{2} - \Fluxintk{1}}{\Lzero(\Coerck{2} - \Coerck{1})}, \quad \Weightk{k} = \frac{\Fluxintk{k+1} - \Fluxintk{k}}{\Lzero(\Coerck{k+1} - \Coerck{k})} - \sum_{i=1}^{k-1} \Weightk{i},
\end{aligned}
\end{equation} 
based on the piecewise nature of the ROHF model approximation, imposing exact internal flux matches of reference and approximation in the already defined coercivity points. In Eq.~\eqref{eqn:recursion}, $\Fluxintk{k}$ indicates the observed internal flux of the reference simulation at transport current level $\I = \Coerck{k}$ within the initial transport current ramp up from virgin state, with an additional coercivity $\Coerck{\Cellnum+1} = \max(\I)$. The initial value for the $\Cellnum$ time constants $\Timeck{k}$ is set to low values (e.g., $0.1\,\si{\micro\second}$) in order to significantly undershoot the physical relaxation behavior for the eddy current development in the strand. This ensures a consistent start with an effectively rate-independent chain of cells. The proposed strategy for the initial parameter guesses, in combination with standard optimizers, typically yields relative flux errors lower than $10\%$ for wide ranges of current amplitude and rate of change.

\section{Results} 

\begin{table}[ht]
\renewcommand{\arraystretch}{1.2}
\centering
\caption{ROHF model parameters for the superconducting composite strand shown in Fig.~\ref{fig:strand_3D} using a chain of $\Cellnum=7$ rate-dependent cells.}
\label{tab:rohf_params}
\begin{tabular}{l||lll}
$k$ & $\Coerck{k}$ ($\si{\ampere}$) & $\Weightk{k}$ & $\Timeck{k}$ ($\si{\milli\second}$) \\
\hline
1 & 6.244 & 0.36 & 0.0\\
2 & 376.8 & 0.08 & 0.47\\
3 & 802.0 & 0.16 & 0.31\\
4 & 1292 & 0.20 & 0.31\\
5 & 1673 & 0.55 & 0.26\\
6 & 2280 & 1.06 & 0.21\\
7 & 2794 & 3.55 & 9.15
\end{tabular}
\end{table}

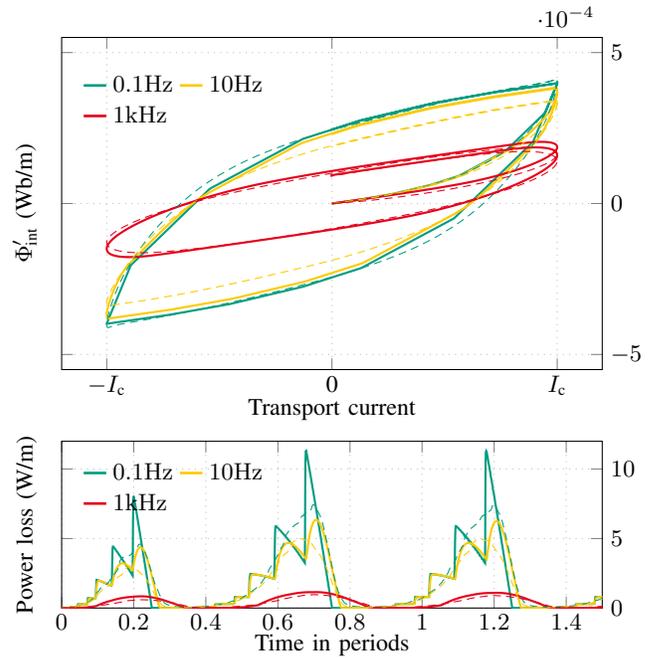
\begin{figure}[ht!]
\centering
\begin{subfigure}[t]{0.99\linewidth}  
\centering
\begin{tikzpicture}[trim axis right, trim axis left][font=\small]
\pgfplotsset{set layers}
 	\begin{axis}[
	tick label style={/pgf/number format/fixed},
    width=\linewidth,
    height=6.0cm,
    ylabel={$\Fluxint$ (Wb/m)},
    ytick={yshift=-0.5em},
    yticklabel style={xshift=0em},
    axis line style={draw=none},
    tick style={draw=none},
    xticklabel=\empty,
    yticklabel=\empty,
    yticklabel pos=left,
    ]
    \end{axis}
 	\begin{axis}[
	tick label style={/pgf/number format/fixed},
    width=\linewidth,
    height=6.0cm,
    grid = major,
    grid style = dotted,
    ymin=-5.5e-4, 
    ymax=5.5e-4,
    xtick={-2960, 0, 2960},
    xticklabels={$-\Ic$, 0 , $\Ic$},
	xlabel={Transport current},
    ylabel style={yshift=-0.5em},
    xlabel style={yshift=0.5em},
    xticklabel style={yshift=0.1em},
    yticklabel pos=right,
    legend columns=2,
    legend style={at={(0.21, 0.7)}, cells={anchor=west}, anchor=south, draw=none,fill opacity=0, text opacity = 1, legend image code/.code={\draw[##1,line width=1pt] plot coordinates {(0cm,0cm) (0.3cm,0cm)};}}
    ]
    \addplot[tu_3, thick] 
    table[x=ROHF_f0.1_IIC1.0_x,y=ROHF_f0.1_IIC1.0_y]{data/current_flux_ratedep.txt};
    \addplot[tu_6, thick] 
    table[x=ROHF_f10_IIC1.0_x,y=ROHF_f10_IIC1.0_y]{data/current_flux_ratedep.txt};
    \addplot[tu_9, thick] 
    table[x=ROHF_f1000_IIC1.0_x,y=ROHF_f1000_IIC1.0_y]{data/current_flux_ratedep.txt};
    \legend{$0.1 \si{\hertz}$, $10 \si{\hertz}$, $1 \si{\kilo\hertz}$};
    \addplot[tu_3, densely dashed] 
    table[x=f0.1_IIC1.0_x,y=f0.1_IIC1.0_y]{data/current_flux_ratedep.txt};
    \addplot[tu_6, densely dashed] 
    table[x=f10_IIC1.0_x,y=f10_IIC1.0_y]{data/current_flux_ratedep.txt};
    \addplot[tu_9, densely dashed] 
    table[x=f1000_IIC1.0_x,y=f1000_IIC1.0_y]{data/current_flux_ratedep.txt};
    \end{axis}
\end{tikzpicture}
\end{subfigure}
\vspace{0.01cm} 
\begin{subfigure}[t]{0.99\linewidth}  
\centering
\begin{tikzpicture}[trim axis right, trim axis left][font=\small]
\pgfplotsset{set layers}
 	\begin{axis}[
	tick label style={/pgf/number format/fixed},
    width=\linewidth,
    height=3.8cm,
    ylabel={Power loss (W/m)},
    ytick={yshift=-0.5em},
    yticklabel style={xshift=0em},
    axis line style={draw=none},
    tick style={draw=none},
    xticklabel=\empty,
    yticklabel=\empty,
    yticklabel pos=left,
    ]
    \end{axis}
 	\begin{axis}[
	tick label style={/pgf/number format/fixed},
    width=\linewidth,
    height=3.8cm,
    grid = major,
    grid style = dotted,
    ymax=12,
    ymin=0,
    xmax=1.5,
    xmin=0,
	xlabel={Time in periods},
    xlabel style={yshift=0.5em},
    xticklabel style={yshift=0.1em},
    yticklabel style={xshift=0em},
    yticklabel pos=right,
    legend columns=2,
    legend style={at={(0.21, 0.5)}, cells={anchor=west}, anchor=south, draw=none,fill opacity=0, text opacity = 1, legend image code/.code={\draw[##1,line width=1pt] plot coordinates {(0cm,0cm) (0.3cm,0cm)};}}
    ]
    \addplot[tu_3, thick] 
    table[x=ROHF_f0.1_IIC1.0_x,y=ROHF_f0.1_IIC1.0_y]{data/time_dynloss_ratedep.txt};
    \addplot[tu_6, thick] 
    table[x=ROHF_f10_IIC1.0_x,y=ROHF_f10_IIC1.0_y]{data/time_dynloss_ratedep.txt};
    \addplot[tu_9, thick] 
    table[x=ROHF_f1000_IIC1.0_x,y=ROHF_f1000_IIC1.0_y]{data/time_dynloss_ratedep.txt};
    \legend{$0.1 \si{\hertz}$, $10 \si{\hertz}$, $1 \si{\kilo\hertz}$};
    \addplot[tu_3, densely dashed] 
    table[x=f0.1_IIC1.0_x,y=f0.1_IIC1.0_y]{data/time_dynloss_ratedep.txt};
    \addplot[tu_6, densely dashed] 
    table[x=f10_IIC1.0_x,y=f10_IIC1.0_y]{data/time_dynloss_ratedep.txt};
    \addplot[tu_9, densely dashed] 
    table[x=f1000_IIC1.0_x,y=f1000_IIC1.0_y]{data/time_dynloss_ratedep.txt};
    \end{axis}
\end{tikzpicture}
\end{subfigure}
\caption{Internal flux hysteresis and related total power loss approximation, Eq.~\eqref{eqn:Ptot}, of a fitted ROHF model with parametrization according to table~\ref{tab:rohf_params} (solid lines) compared to FE references (dashed lines) for three frequencies of sinusoidal transport current with critical current amplitude.}
\vspace{-0.5cm}
\label{fig:hysteresis_rohf}
\end{figure}

In a first verification step, the internal magnetic flux is compared between a reference FE CATI model of the 54-filament strand illustrated in Fig.~\ref{fig:strand_3D}, with the geometric specifications according to~\cite{dular24cati}, and a fitted chain of $\Cellnum=7$ rate-dependent ROHF cells with parametrization according to Tab.~\ref{tab:rohf_params}, taken from~\cite{glock25}. The initial state for both computations is a virgin strand. Applying one and a half periods of sinusoidal transport current with the critical strand current $\Ic=2960\si{\ampere}$ as amplitude at three frequencies yields the internal flux hysteresis approximations shown in Fig.~\ref{fig:hysteresis_rohf} - top.

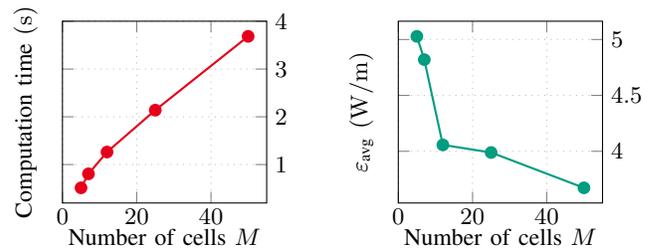
\begin{figure}[ht]
\centering
\begin{subfigure}[t]{0.49\linewidth}  
\begin{tikzpicture}[trim axis right][font=\small]
\pgfplotsset{set layers}
 	\begin{axis}[
	tick label style={/pgf/number format/fixed},
    width=0.99\linewidth,
    height=4cm,
    ylabel={Computation time $\left(\si{\second}\right)$},
    ytick={yshift=-0.5em},
    yticklabel style={xshift=0em},
    axis line style={draw=none},
    tick style={draw=none},
    xticklabel=\empty,
    yticklabel=\empty,
    yticklabel pos=left,
    ]
    \end{axis}
 	\begin{axis}[
	tick label style={/pgf/number format/fixed},
    width=0.99\linewidth,
    height=4cm,
    grid = major,
    grid style = dotted,
    xmin=0, xmax=55,
    xlabel={Number of cells $\Cellnum$},
    xlabel style={yshift=0.5em},
    xticklabel style={yshift=0.1em},
    yticklabel style={xshift=0em},
    yticklabel pos=right,
    legend columns=1,
    legend style={at={(0.02, 0.7)}, cells={anchor=west}, anchor=west, draw=none,fill opacity=0, text opacity = 1}
    ]    
    \addplot[color=tu_9, thick, mark=*] 
    table[x=cells,y=time]{data/cells_error_time.txt};
    \end{axis}
\end{tikzpicture}
\end{subfigure}
\centering
\begin{subfigure}[t]{0.49\linewidth}  
\begin{tikzpicture}[trim axis right][font=\small]
\pgfplotsset{set layers}
 	\begin{axis}[
	tick label style={/pgf/number format/fixed},
    width=0.99\linewidth,
    height=4cm,
    ylabel={$\varepsilon_\text{avg}$ $(\si{\watt/ \meter})$},
    ytick={yshift=-0.5em},
    yticklabel style={xshift=0em},
    axis line style={draw=none},
    tick style={draw=none},
    xticklabel=\empty,
    yticklabel=\empty,
    yticklabel pos=left,
    ]
    \end{axis}
 	\begin{axis}[
	tick label style={/pgf/number format/fixed},
    width=.99\linewidth,
    height=4cm,
    grid = major,
    grid style = dotted,
    xmin=0, xmax=55,
    xlabel={Number of cells $\Cellnum$},
    xlabel style={yshift=0.5em},
    xticklabel style={yshift=0.1em},
    yticklabel style={xshift=0em},
    yticklabel pos=right,
    legend columns=3,
    legend style={at={(0, 0.85)}, cells={anchor=west}, anchor=west, draw=none,fill opacity=0, text opacity = 1}
    ]    
    \addplot[color=tu_3, thick, mark=*] 
    table[x=cells,y=error]{data/cells_error_time.txt};
    \end{axis}
\end{tikzpicture}
\end{subfigure}
\caption{Impact of cell number $\Cellnum$ on the computation time for a benchmark case of ten thousand time steps (left) and the mean absolute power loss error $\varepsilon_\text{avg}$ of the approximation (right) for a sinusoidal transport current signal with amplitude $\Ic$ and frequency of $10~\si{\hertz}$.}
\label{fig:cellcount}
\end{figure}

Noticeable deviations between the ROHF approximations and the FE references appear in the low-frequency regime, where eddy currents are negligible and the piecewise linear description of the internal flux by the ROHF model exhibits sharp transitions at the coercivity values of each cell (Fig.~\ref{fig:hysteresis_rohf} - top). These discontinuities in the hysteretic flux, and the resulting jumps in its time derivative at low transport current frequencies, lead to fluctuations in the total power loss prediction of the ROHF chain Eq.~\eqref{eqn:Ptot} around the reference values (Fig.~\ref{fig:hysteresis_rohf} - bottom). This error can be reduced by increasing the number of reduced-order cells, which allow for a better approximation of the flux with less pronounced discontinuities, at the cost of increased computational effort, as illustrated in Fig.~\ref{fig:cellcount}.

Typical computation times for the FE references with detailed current density resolution within the strand lie in the range of one to three hours, while the evaluation of the corresponding chain of ROHF cells is done within one second, leading to a gain of several orders of magnitude in computational efficiency. In a straightforward implementation as a standalone algorithm, this processing time of the ROHF model scales linearly with the number of reduced-order cells $\Cellnum$ (Fig.\ref{fig:cellcount} - left). This is because of the $\Cellnum$ individual cell updates, Eq.\eqref{eqn:scalar_play_eddy}, performed in sequence. However, since all cells are driven by the same change in transport current, it is possible to parallelize parts of this evaluation, which could significantly reduce the observed scaling factor. Conversely, when integrating this algorithm into a FE code, the scaling of the sequential evaluation will increase beyond linear. This is due to the current split per cell defined in Eq.~\eqref{eqn:current_split_eddy}, which introduces at least $\Cellnum$ additional degrees of freedom into the solver scheme at every time step.

The absolute mean error in the power loss approximation of the ROHF model, used as a measure of the deviations observed in Fig.~\ref{fig:hysteresis_rohf}, decreases with increasing cell numbers (Fig.~\ref{fig:cellcount} - right). While a higher number of cells yields smoother and slightly more accurate power loss approximations, using longer chains of ROHF cells (i.e., $\Cellnum \gtrsim 10$) is not necessary for homogenizing transport current effects in the fine structures of accelerator magnet simulations. Indeed, the predicted power loss serves as a heat source in a thermal diffusion problem, which inherently smooths rapid variations. As a result, even a relatively small number of cells can yield sufficiently accurate approximations. Furthermore, the accuracy of the ROHF model is inherently limited by its underlying assumptions. The ROHF model simplifies the complex dynamics of a superconducting strand by describing all geometric effects with lumped quantities. Consequently, a small residual approximation error always remains compared to reference FE solutions.

\section{Conclusion}  

The presented ROHF model can be tuned to replicate the macroscopic response of superconducting strands to transport current variations without resolving the internal current density distribution, and is thus capable of reducing computation times by several orders of magnitude while offering a very good accuracy. The actual speed-up factor compared to detailed models depends on the chosen number of reduced-order cells $\Cellnum$, which can be adjusted to balance approximation accuracy and computational efficiency. Due to the one-time \textit{a priori} identification of the $3\Cellnum$ strand-dependent model parameters from reference simulations, the method is especially well-suited for recurring modeling scenarios that involve a limited number of strand types, where its efficiency can be fully leveraged.

In future developments, the ROHF model will be integrated into superconducting magnet FE models via field-circuit coupling approach. To this end, it will be combined with the Reduced Order Hysteretic Magnetization (ROHM) model~\cite{dular25rohm} in order to capture the strand dynamics, including inductance and magnetization, without resolving their fine structures.

\section*{Acknowledgment}  

This work was partially supported by the CERN High Field Magnets (HFM) program.

\ifCLASSOPTIONcaptionsoff
  \newpage
\fi

\end{document}